\documentclass{article}
\usepackage[T1]{fontenc}
\usepackage[utf8]{inputenc}
\usepackage{prettyref}
\usepackage{float}
\usepackage{amstext}
\usepackage{graphicx}
\usepackage[numbers]{natbib}

\makeatletter
\usepackage{prettyref}
\newrefformat{eq}{eqn\nolinebreak\,\nolinebreak(\ref{#1})}
\newrefformat{sec}{Section\nolinebreak\,\nolinebreak\ref{#1}}
\newrefformat{cha}{Chapter\nolinebreak\,\nolinebreak\ref{#1}}
\newrefformat{tab}{Table\nolinebreak\,\nolinebreak\ref{#1}}
\newrefformat{fig}{Figure\nolinebreak\,\nolinebreak\ref{#1}}
\newrefformat{lem}{Lemma\nolinebreak\,\nolinebreak\ref{#1}}
\newrefformat{thm}{Theorem\nolinebreak\,\nolinebreak\ref{#1}}
\newrefformat{axm}{Axiom\nolinebreak\,\nolinebreak\ref{#1}}
\newrefformat{app}{Appendix\nolinebreak\,\nolinebreak\ref{#1}}
\newrefformat{exe}{Exercise\nolinebreak\,\nolinebreak\ref{#1}}
\newrefformat{cor}{Corollary\nolinebreak\,\nolinebreak\ref{#1}}
\newrefformat{pro}{Property\nolinebreak\,\nolinebreak\ref{#1}}
\newrefformat{def}{Definition\nolinebreak\,\nolinebreak\ref{#1}}
\newrefformat{subsec}{Section\nolinebreak\,\nolinebreak\ref{#1}}
\usepackage{amsmath}
\usepackage{bm}
\numberwithin{equation}{section}
\numberwithin{table}{section}
\newcommand{\br}[1]{\mathbf{#1}} %bold upright roman
\newcommand{\bg}[1]{\bm{#1}} %bold greek
\usepackage{pifont}
\usepackage{txfonts}
\usepackage{charter}
\date{}

\makeatother

\begin{document}

\title{A Spherical Version of Feynman’s Static Field Momentum Example}

\author{{\normalsize{}Oliver Davis Johns}\\
{\normalsize{}San Francisco State University, Physics and Astronomy
Department}\\
{\normalsize{}1600 Holloway Avenue, San Francisco, CA 94133, USA}\\
{\normalsize{}Email: ojohns@metacosmos.org}\\
{\normalsize{}Web: https://www.metacosmos.org}}
\maketitle
\begin{abstract}
The Feynman demonstration that electromagnetic field momentum is real—even
for static fields—can be made more pedagogically useful by simplifying
its geometry. Instead of Feynman's disk with charged balls on its
surface, this article uses the geometry of a hollow non-conducting
sphere with uniform surface charge density. With only methods available
in a typical upper-division electrodynamics course, the initial field
angular momentum and the final mechanical angular momentum can then
be calculated in closed form and shown to be equal.

This spherical geometry also provides a counterexample to the idea
that electromagnetic field momentum is due to the classical flow of
an inertial relativistic mass defined as the energy density divided
by the square of the speed of light. The curved flow lines of an inertial
field momentum would require a centripetal force to bend them, but
no such force can be identified classically.
\end{abstract}

\section{Introduction}

\label{sec:intro}The Feynman lectures\footnote{Feynman et al \citep{feynmanlectures}, Section 17-4, Section 27-6,
and Figure 17-5.} give an example of the reality of field momentum in a static electromagnetic
field. A non-conducting, non-magnetic, horizontal disk is free to
rotate about a vertical axis through its center. Several identical,
charged balls are equally spaced around its circumference. In its
center is an electromagnet. Due to cancellation, the net electric
field is small on the surface of the disk. For distances from the
axis of the disk larger than its radius, the electric and magnetic
fields give a Poynting vector with a significant azimuthal component.
The angular momentum density associated with this Poynting vector
produces a nonzero total field angular momentum.
\begin{figure}[H]
\centerline{\includegraphics{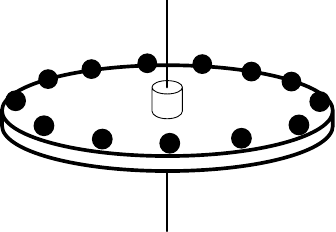}}\caption{\footnotesize{The Feynman example. Charged balls are spaced evenly
around the circumference of a disk. A small electromagnet is placed
at the center.}}
\end{figure}
The disk is at rest for times less than zero and hence has no mechanical
angular momentum. If the electromagnet is programmed to turn itself
off beginning at time zero, the resulting induced electric field exerts
a torque on the charged spheres. After a long time, the magnetic field,
and hence the Poynting vector and the field angular momentum, are
zero. But the disk is now rotating as a result of the torque exerted
on the spheres during the decay of the magnetic field. The initial
field angular momentum has become final mechanical angular momentum,
which shows that the initial field angular momentum must have been
real.

The difficulty with this example is that it depends on qualitative
arguments about the field strengths. Obviously, the magnetic field
is stronger closer to the electromagnet. One must argue that the weakness
of the electric field there more than compensates, so that the Poynting
vector is large only in the region outside the radius of the disk
where its azimuthal sense is correct. Such qualitative argumentation
is not helpful for students who are, after all, not totally convinced
that the phenomenon of field angular momentum is real.

Fortunately, it is not difficult to concoct a simplified, spherical
example with fewer ambiguities. The methods used: multipole expansion
of magnetic field and magnetic vector potential, Gauss’s law for electric
fields, retarded potentials for time-varying currents, the expression
for the electric field when vector potential $\br A$ varies, are
all part of a standard junior level Electricity and Magnetism course.\footnote{For example, Wangsness \citep{wangsness} and Griffiths \citep{griffiths}.}
That level is also ideal for the introduction of examples such as
this one. 

\section{The Spherical Feynman Example: \protect \\
Static-Field Angular Momentum}

\label{sec:static}Suppose that a non-conducting, non-magnetic spherical
shell of outer radius $a$ is suspended by a vertical axis through
its center (taken as the $z$-axis), about which it is free to rotate.
The surface at radius $a$ has a fixed, uniform surface charge density
$\sigma_{0}$. Gauss's law then shows the electric field to be zero
for $r<a$. At the center of the sphere is an electromagnet configured
so that its total vector magnetic moment is $\eta_{0}\hat{\br z}$,
parallel to the vertical axis. As in the original Feynman example,
the magnetic moment $\eta_{0}$ is a positive constant for $t<0$,
and decreases to zero for $t>0$. The sphere is at rest for negative
time; there is nothing moving and hence no mechanical momentum for
$t<0$.
\begin{figure}[H]
\centerline{\includegraphics[scale=0.5]{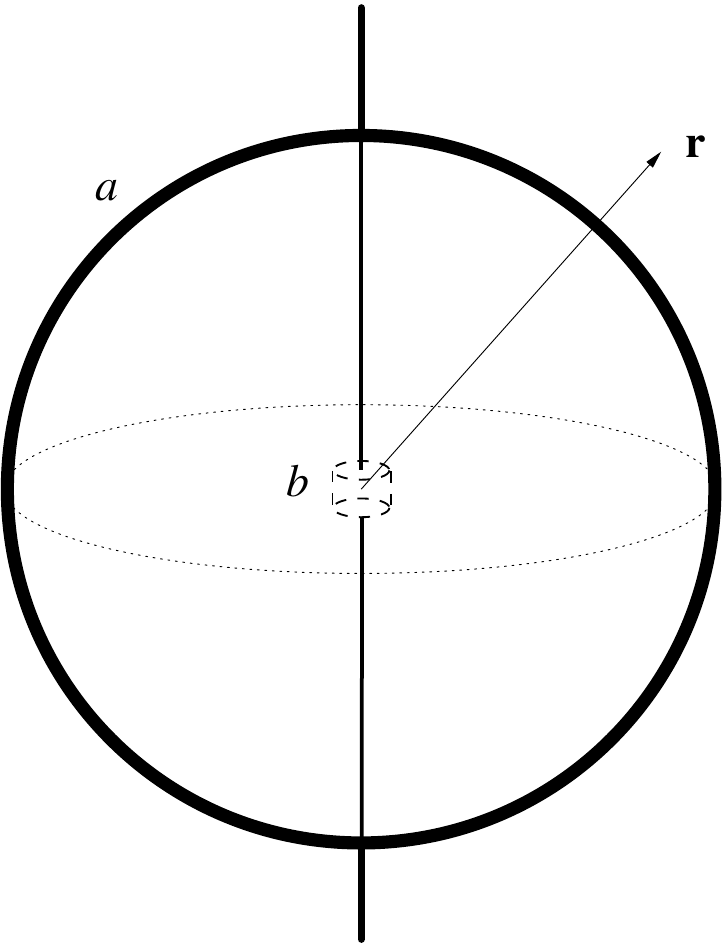}}

\caption{\label{fig:2}\footnotesize{The simplified Feynman example. A non-conducting
spherical shell of outer radius $a$ is free to rotate about a fixed,
non-conducting, and non-magnetic vertical shaft. At the center of
the sphere, a small electromagnet is attached to the shaft.} }
\end{figure}

All currents in the electromagnet are confined to be less than distance
$b$ from the sphere’s center. Take $b/a$ to be sufficiently small
that, in the region on and outside radius $a$ where the electric
field is nonzero, the magnetic field can be approximated adequately
by just the dipole term in a multipole expansion. The electric field
is zero for $r<a$ and is given by Gauss's law for $r>a$. Thus the
static fields for $r>a$ are\footnote{This paper uses MKSA units in vacuum with the speed of light $\left(\varepsilon_{0}\mu_{0}\right)^{-1/2}$
denoted by $c$. Spherical polar coordinates $r,\theta,\phi$ are
used with unit vectors $\hat{\br r}$, $\hat{\bg\theta}$, $\hat{\bg\phi}$,
the $z$-axis vertical along the shaft, and the origin at the center
of the sphere. The $x$ and $y$-axes are not shown in \prettyref{fig:2}.}
\begin{equation}
\br E=\dfrac{a^{2}\sigma_{0}}{\varepsilon_{0}r^{2}}\:\hat{\br r}\quad\quad\quad\quad\br B=\dfrac{\mu_{0}\eta_{0}}{4\pi r^{3}}\left(2\cos\theta\:\hat{\br r}+\sin\theta\:\hat{\bg\theta}\right)\label{eq:2.1}
\end{equation}
The Poynting vector is $\br S=\br E\times\br B/\mu_{0}$ and the momentum
density is $\br G=\br S/c^{2}$. The momentum density is thus zero
for $r<a$ and 
\begin{equation}
\br G=\frac{\mu_{0}a^{2}\sigma_{0}\eta_{0}\sin\theta}{4\pi r^{5}}\:\hat{\bg\phi}\label{eq:2.2}
\end{equation}
for $r>a$. The field angular momentum density for $r>a$ is therefore
\begin{equation}
\br r\times\br G=-\frac{\mu_{0}a^{2}\sigma_{0}\eta_{0}\sin\theta}{4\pi r^{4}}\:\hat{\bg\theta}\label{eq:2.3}
\end{equation}
The total initial static field angular momentum is then
\begin{equation}
\br L_{0}=\int_{a}^{\infty}\!\!\!dr\:r^{2}\int_{0}^{\pi}\!\!\!d\theta\sin\theta\int_{0}^{2\pi}\!\!\!d\phi\:\br r\times\br G=\dfrac{2\mu_{0}a\sigma_{0}\eta_{0}}{3}\:\hat{\br z}\label{eq:2.4}
\end{equation}

\section{The Spherical Feynman Example: \protect \\
Torques and their Impulse}

\label{sec:torque}At time zero, the current in the electromagnet
begins to decrease slowly. Take its magnetic moment for $t\ge0$ to
be $\eta(t)=\eta_{0}\text{e}^{-t/\tau}$. Taking $\tau\gg a/c$ so
that time delay effects are negligible, there is an instantaneous
magnetic vector potential in the vicinity of $r=a$ given, again,
by just the dipole term in the multipole expansion
\begin{equation}
\br A(\br r,t)=\dfrac{\mu_{0}\eta(t)\sin\theta}{4\pi r^{2}}\:\hat{{\bf \phi}}\label{eq:3.1}
\end{equation}
From this vector potential and the surface charge density $\sigma_{0}$,
we use $\br E=-\nabla\Phi_{\text{elec}}-\partial\br A/\partial t$
to derive the instantaneous electric field at radius $r=a$
\begin{equation}
\br E(a,\theta,\phi,t)=\dfrac{\sigma_{0}}{\varepsilon_{0}}\:\hat{\br r}-\dfrac{\mu_{0}\sin\theta}{4\pi a^{2}}\,\dfrac{d\eta}{dt}\:\hat{\bg\phi}\label{eq:3.2}
\end{equation}
The total instantaneous torque of this electric field on the surface
charge density is then
\begin{equation}
\br N=a^{2}\int_{0}^{\pi}\!\!\!d\theta\sin\theta\int_{0}^{2\pi}\!\!\!d\phi\:a\:\hat{\br r}\times\sigma_{0}\,\br E(a,\theta,\phi,t)=-\dfrac{2\mu_{0}a\sigma_{0}}{3}\,\dfrac{d\eta}{dt}\:\hat{\br z}\label{eq:3.3}
\end{equation}
The total impulse of this torque is the final, mechanical angular
momentum
\begin{equation}
\br L_{\text{f}}=\int_{0}^{\infty}\!\!\!dt\,\br N=\dfrac{2\mu_{0}a\sigma_{0}\eta_{0}}{3}\:\hat{\br z}\label{eq:3.4}
\end{equation}

The initial field angular momentum \prettyref{eq:2.4} equals the
final mechanical angular momentum \prettyref{eq:3.4}. The simplifying
assumptions made are internally consistent. We explicitly assumed
$a\gg b$, and $\tau\gg a/c$. An implicit assumption, that the magnetic
field produced by the rotating sphere produces negligible residual
field angular momentum, can be ensured by using a sphere with very
large moment of inertia so that its angular velocity remains small.

\section{The Quantitative Feynman Argument}

The principle of angular momentum conservation requires that the final
mechanical angular momentum $\br L_{\text{f}}$ of the rotating sphere
at $\tau\gg a/c$ must have already been present for $t<0$. Since
there was no motion for $t<0$, that angular momentum can only have
been stored in the initial static field configuration. 

The quantitative agreement $\br L_{\text{f}}=\br L_{0}$ between \prettyref{eq:3.4}
and \prettyref{eq:2.4} is convincing proof that the final mechanical
angular momentum was indeed stored initially in the static electromagnetic
field, and that the static field angular momentum $\br L_{0}$ must
therefore have been physically real. It follows that the linear momentum
$\br G$ in \prettyref{eq:2.2} must also be physically real.\footnote{The $\br G$ in \prettyref{eq:2.2} has no component parallel to $\hat{r}$
and hence the whole of momentum density $\br G$ contributes to the
angular momentum density $\br r\times\br G$ in \prettyref{eq:2.3}.}

\section{Other Geometries\label{sec:Other}}

Previous attempts to verify the Feynman example quantitatively have
used geometries not closely related to the Feynman disk with balls
on its surface. 

Romer \citep{romer-angmom} uses a very long, hollow, fixed solenoid
containing a fixed, coaxial conducting rod. At $t=0$, a spot of radioactive
material on the rod emits a particle of positive charge which pierces
the solenoid and escapes to infinity. As it escapes, this particle
is given some angular momentum by the magnetic field of the solenoid.
After the particle escapes to infinity, there is field angular momentum
from the combined electromagnetic fields of the solenoid and the now-charged
rod. This field angular momentum is shown to be equal in magnitude
and opposite in direction to the angular momentum that was conveyed
to the escaping particle. However, this example, and a second one
in Romer's paper, do not directly show field angular momentum becoming
mechanical angular momentum.

Lombardi \citep{Lombardi-Feyn} provides a formal and geometry-independent
proof that angular momentum transferred to charges by a changing magnetic
field must equal the change in field angular momentum. However this
paper does not calculate specific values in a pedagogically useful
way.

Corinaldesi \citep{corinaldesi-feyn} and Boos \citep{boos-angmom}
also use a long, fixed, hollow solenoid. It contains two hollow, coaxial,
nonconducting cylinders that have opposite surface charge densities
and are free to rotate about their common axis. This geometry is like
the Feynman example inside-out. The electromagnet is a solenoid on
the outside and the moving parts are inside. These papers then turn
off the magnetic field slowly and calculate the total impulse of the
torque on the inner cylinders, showing it to equal the initial field
angular momentum. However, this geometry can only compare values per
unit length of the cylinder, and also has problems with fringing at
the cylinders' top and bottom.

\section{Is Field Momentum Due to Classical\protect \\
Flow of Inertial Mass?\label{sec:MassFlow}}

The motivation for the spherical Feynman geometry in this paper is
principally pedagogical—to show students the reality of static field
momentum. However, a curious student will then wonder how a dynamic
thing like momentum can possibly be contained in a static field. The
spherical Feynman geometry is an ideal testbed to consider some answers
to that question.

An entirely classical explanation of static field momentum has been
proposed,\footnote{See Sebens \citep{sebens-mass}.} based on the
idea that electromagnetic energy ${\cal E}=\left(\varepsilon_{0}E^{2}+B^{2}/\mu_{0}\right)/2$
contains an inertial relativistic mass ${\cal M}={\cal E}/c^{2}$.
Then a conventional formula defining the velocity $\br v$ of energy
flow 
\begin{equation}
\br S={\cal E}\,\br v\label{eq:5.1a}
\end{equation}
can be divided by $c^{2}$ to yield a relation between momentum $\br G=\br S/c^{2}$
and relativistic mass
\begin{equation}
\br G=\dfrac{\br S}{c^{2}}=\dfrac{{\cal E}}{c^{2}}\,\br v={\cal M}\,\br v\label{eq:5.1b}
\end{equation}
thus exhibiting static field momentum as due to the flow of inertial
relativistic mass. 

But, in spite of its plausibility, the energy-flow velocity $\br v$
defined in \prettyref{eq:5.1a} and used in \prettyref{eq:5.1b} is
inconsistent with the transformation rules of special relativity.
There is no relativistically correct velocity $\br v$ that satisfies
\prettyref{eq:5.1a} and \prettyref{eq:5.1b}.\footnote{See Propositions 1 and 2 of the 2023 revision of Johns \citep{oj-energyflow}.}
The field momentum $\br G$ cannot be explained classically as due
to the flow of an inertial relativistic mass ${\cal M={\cal E}}/c^{2}$.

The spherical Feynman example also permits a more direct argument
against a classical explanation based on an inertial mass ${\cal M}$.
Evaluating \prettyref{eq:2.2} in the equatorial plane $\theta=\pi/2$
at $r=2a$ shows the field lines of momentum $\br G(r,\theta,\phi)$
there as forming a circle of radius $2a$ 
\begin{equation}
\br G(2a,\pi/2,\phi)=\frac{\mu_{0}\sigma\eta_{0}}{32\pi a^{3}}\:\hat{\bg\phi}\label{eq:6.1}
\end{equation}
If there is some velocity $\br v$ and inertial mass ${\cal M}$ such
that ${\cal M}\,\br v=\br G$, then a centripetal force density 
\begin{equation}
\br F=-\dfrac{{\cal M}\,v^{2}}{2a}\hat{r}\label{eq:6.2}
\end{equation}
is required to hold that mass in its circular orbit. 

But there is no such force density $\br F$ in classical electromagnetism.
Charges and currents act on fields, and fields act on charges and
currents, but fields do not act on fields. Classical electrodynamics
is a linear theory. 

The spherical Feynman example is thus an ideal vehicle for demonstrating
to students both that static field momentum is real, and also that
it cannot be explained as the classical flow of some kind of inertial
mass.

\end{document}